# SINGLE STRANDED DNA TRANSLOCATION THROUGH A FLUCTUATING NANOPORE

**O. Flomenbom and J. Klafter**
School of Chemistry, Raymond & Beverly Sackler Faculty of Exact Sciences,
Tel-Aviv University, Tel-Aviv 69978, Israel

ABSTRACT

We investigate the translocation of a single stranded DNA (ssDNA) through a pore, which fluctuates between two conformations, by using coupled master equations (ME). The probability density function (PDF) of the first passage times (FPT) of the translocation process is calculated, displaying a triple, double or mono-peaked behavior, depending on the system parameters. An analytical expression for the mean first passage time (MFPT) of the translocation process is derived, and provides an extensive characterization of the translocation process.

## INTRODUCTION

Translocation of biopolymers through a membrane pore occurs in a variety of biological processes, such as gene expression in eucaryotic cells [1], conjugation between procaryotic cells, and virus infection [2]. The importance of translocation in biological systems and its possible applications have been the motivation for recent theoretical and experimental work on this topic. In experiments one usually measures the time it takes a single voltage-driven ssDNA to translocate through an $\alpha$−hemolysin channel of a known structure [3]; see Fig. 1 for an illustration of the process. Since ssDNA is negatively charged (each monomer of length $b$ has an effective charge of $zq$, where $q$ is the electron charge, and $z$ (0<$z$<1) is controlled by the solution pH and strength), when applying a voltage the polymer is subject to a driving force while passing through the transmembrane pore part (TPP) from the negative (cis) side to the positive (trans) side. Because the presence of the ssDNA in the TPP blocks the cross-TPP current, one can deduce the FPT PDF, $F(t)$, from the current blockade duration times [4,5].

Experiments by Kasianowicz et al. [4], show $F(t)$ with three peaks. It was suggested that the short-time peak represents the non-translocated events, while the other two longer-time peaks represent translocation events of different ssDNA orientations. In addition, the times that maximize the translocation peaks were shown to be proportional to the polymer length and inversely proportional to the applied field. In experiments by Meller et al., $F(t)$ was shown to be mono-peaked, with a corresponding maximizing time that has inverse quadratic field dependence.

The theoretical models used to describe the translocation are one-dimensional with the structure of the pore taken to be rigid; namely, the structure is governed by a single conformation [6-10]. Here we relax the assumption of a single pore conformation and introduce a second conformation coupled to the first one. Using this generalized model we calculate $F(t)$ and the MFPT, which provide an extensive characterization of the translocation process. Our model, we believe, helps gain insight into the translocation of a polymer through a narrow pore, and provides an explanation for the diversity of experimental observations [4-5].

**THE MODEL**

We first describe our basic model for the translocation process [6]. We map the three-dimensional translocation process onto a discrete one-dimensional space containing $n(=N+d-1)$ states separated from each other by a unit length $b$, and use an $n$-state ME for describing the translocation of an $N$-monomer long ssDNA subject to an external voltage $V$ and temperature $T$. The translocation takes place within a TPP of a length that corresponds to $d$ monomers. The occupation PDF of the $j$ state is $[\vec{P}(t)]_j = P_j(t)$, where the state index $j$ determines the number of monomers on each side of the membrane and within the TPP ($m_j$). $P_j(t)$ satisfies the equation of motion

$$\partial P_j(t)/\partial t = a_{j+1,j} P_{j+1}(t) + a_{j-1,j} P_{j-1}(t) - (a_{j,j+1} + a_{j,j-1}) P_j(t) \tag{1}$$

under absorbing boundary conditions on both sides of the membrane (the polymer can exit the TPP on both sides). Eq. 1 can be written in a matrix representation, $\partial \vec{P}(t)/\partial t = \mathbf{A}\vec{P}(t)$, where the propagation matrix $\mathbf{A}$ is a tridiagonal matrix that contains information about the transitions between states in terms of rate constants, $a_{j,j\pm1}$, which are given by: $a_{j,j\pm1} = k_j p_{j,j\pm1}$. Here $k_j$ is the rate to perform a step, $p_{j,j-1}$ ($p_{j,j+1}$) is the probability to move one state from state $j$ to the trans (cis) side, and $p_{j,j-1} + p_{j,j=1} = 1$.

$k_j$ is taken to be similar to the inverse of the longest bulk relaxation time of a polymer [11] $k_j = 1/(\beta \xi_p b^2 m_j^\mu) \equiv R/m_j^\mu$; $\beta^{-1} = k_B T$, with two exceptions: $\xi_p$ represents the ssDNA-TPP interaction and cannot be calculated from the Stokes relation, and $\mu$ serves as a measure of the polymer stiffness inside the confined volume of the TPP, and is bounded by the conventional values [11]: $0 \leq \mu \leq 3/2$. For poly-d$nu$ ($nu$ stands for nucleotide), we estimated that $\xi_p(A_{nu}) \approx 10^{-4} meVs/nm^2$, $\xi_p(C_{nu}) = \xi_p(T_{nu}) = \xi_p(A_{nu})/3$, and $\mu(C_{nu}) = 1$, $\mu(A_{nu}) = 1.14$, $\mu(T_{nu}) = 1.28$, where $A_{nu}$, $C_{nu}$ and $T_{nu}$ stand for adenine, cytosine and thymine nucleotides, respectively.

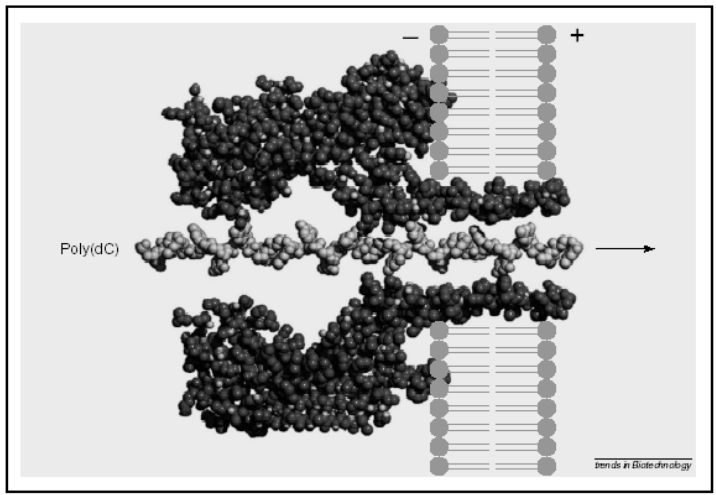

FIG 1 Illustration of the translocation process of a ssDNA (here poly-d$C_{nu}$) through the $\alpha$-hemolysin channel, which is driven by an applied field. The figure was taken from Ref. [12] with the permission of the authors.

Assuming a quasi-equilibrium process, and therefore the detailed balance condition, and using the approximation $a_{j,j-1}/a_{j-1,j} \approx p_{j,j-1}/(1-p_{j,j-1})$, the probability $p_{j,j-1}$ is found to be: $p_{j,j-1} = (1+e^{\beta \Delta E_j})^{-1}$. The free energy difference between states, $\Delta E_j = E_{j-1} - E_j$, is computed considering three contributions: electrostatic, entropic, and an average attractive interaction energy between the ssDNA and the pore. More explicitly, $\beta \Delta E_j$ is given by $\beta \Delta E_j = \beta \Delta E_j^p + \delta_j$ where $\beta \Delta E_j^p \leq 0$ represents the effect of the field which directs towards the trans-side and $\delta_j > 0$ (for $j>d$) represents an effective directionality to the cis-side, which originates from the entropic factors and the average attractive interaction energy between the ssDNA and the pore. From the expressions for $\beta \Delta E_j$ and $p_{j,j-1}$ it follows that the ratio $V/V_C \equiv \beta z |q| V (1+1/d)$ determines the directionality of the translocation, and in particular for $V/V_C > 1$ there is a bias towards the trans-side of the membrane.

A more realistic description of the translocation can be obtained by taking into consideration fluctuations in the structure of the pore. Although it is known that $\alpha$ – hemolysin has a solid structure that allows its crystallization [3], during the translocation of a long polymer through the pore changes in the TPP structure can take place. Accordingly, we introduce an additional pore conformation, which is represented by the propagation matrix $\mathbf{B}$. The changes in the pore conformation between *A* and *B* are controlled by the interconversion rates, $\omega_A$ and $\omega_B$. $\omega_A(\omega_B)$ is the rate of the change from the *A* (*B*) to the *B* (*A*) pore conformation. The physical picture of the process is that when the pore conformation changes, a different environment is created for the ssDNA occupying the TPP. This implies a change in $\xi_p$ and $\mu$. For a large polymer *N>d*, we take $\mathbf{B} \approx \lambda \mathbf{A}$, where $\lambda$ is a (dimensionless) parameter that represents the effect of the conformational change on $\xi_p$ and $\mu$. The parameter $\lambda$ may be interpreted as a measure of an effective available volume in the TPP, when the amino acids residues protruding the TPP change their positions. Understanding of the effect of $\lambda$ on the translocation is achieved by examining several limiting cases. For $\lambda = 0$ movement in any direction occurs only when the ssDNA is subject to the *A* conformation environment. When the ssDNA is subject to the *B* conformation, it is trapped for a period of time governed by the interconversion rates. For $\lambda = 1$, the changes in the pore structure do not affect the translocation, and the process reduces to a translocation through a single pore conformation. A faster translocation relative to the single conformation case is obtained for $\lambda > 1$. In this paper we restrict ourselves to the regime $0 \leq \lambda \leq 1$.

The equations of motion of the ssDNA translocation through the fluctuating pore, written in matrix representation, are:

$$\frac{\partial}{\partial t}\begin{pmatrix} \vec{P}(t;A) \\ \vec{P}(t;B) \end{pmatrix} = \begin{pmatrix} \mathbf{A} - \omega_{\mathbf{A}} & \omega_{\mathbf{B}} \\ \omega_{\mathbf{A}} & \mathbf{B} - \omega_{\mathbf{B}} \end{pmatrix} \begin{pmatrix} \vec{P}(t;A) \\ \vec{P}(t;B) \end{pmatrix} \qquad (2)$$

where $\vec{P}(t;i)$, *i=A, B*, is the occupation PDF vector of configuration *i*, $\omega_\mathbf{i} = \omega_i \mathbf{I}$, and $\mathbf{I}$ is the unit matrix of *n* dimensions.

# THE PDF OF THE TRANSLOCATION TIMES

We turn now to calculate the FPT PDF, which is defined by $F(t) = \partial(1 - S(t))/\partial t$, where $S(t)$ is the survival probability; namely, the probability to have at least one monomer in the pore, which is given by $S(t) = \sum_{i=A}^{B}\sum_{j=1}^{n} P_j(t;i)$. The $P_j(t;i)$'s are obtained by solving Eq. 2 [13]. Fig. 2 shows $F(t)$ for several values of $\lambda$. $F(t)$ is either double or triple-peaked, depending on $\lambda$. The time that maximizes each of the peaks is denoted as $t_{m,i}$ where $i=1,2,3$ (e.g. $t_{m,1}$ is the short-time peak). For the single conformation case we find that $F(t)$ can be either mono or double-peaked depending on $V/V_C$ and on the initial state of the translocation $x$ [6]. The short-time peak represents the non-translocated events, while the long-time peak represents the translocation across the pore. The generalization to two pore conformations yields two translocation peaks in addition to a short-time non-translocation peak. As mentioned, for $\lambda \to 1$ the two conformations reduce to a single one, resulting therefore in one translocation peak. For $\lambda < 1$ the location and amplitude of the slower translocation peak that stems from to the $B$ conformation is determined by the $\lambda$ value, and for $\lambda \to 0$ it spreads out towards larger times, which results in its disappearance. In the inset of Fig. 2 we find that the range of $\lambda$ values for which $F(t)$ exhibits three distinct peaks is $0.1 \leq \lambda \leq 0.3$.

Interestingly enough, as shown in Fig. 3 and in the inset of Fig. 3, $F(t)$ exhibits two peaks corresponding to actual translocation only when $\omega \equiv \omega_A/\omega_B \approx 1$. For $\omega \ll 1$ and $\omega \gg 1$ only one peak corresponding to an actual translocation survives. For all cases there is always a peak representing non-translocation events.

An additional important ratio that determines the $F(t)$ shape is the ratio between the interconversion rate ($\omega_A$ or $\omega_B$) and the dominant rate of the $A$ conformation, which for a sufficiently large $N$ is $k = R/d^\mu$. This ratio gives an estimate of the number of moves done in a given conformation before a change in the pore structure occurs. We find that for two translocation peaks to be obtained, the ratio $\omega_B/k$ (or $\omega_A/k$ due to $\omega \approx 1$) should fulfill $\omega_B/k \leq 10^{-3}$ (data not shown).

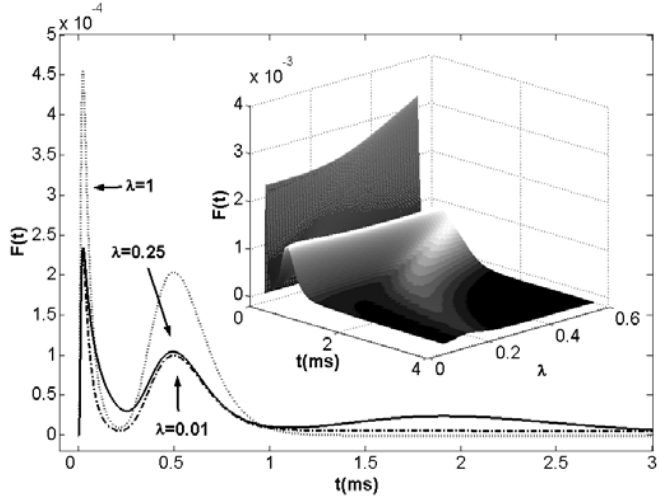

FIG 2 $F(t)$ for poly-$dT_{nu}$ for several values of $\lambda$ with: $N=30$, $d=12$, $x=N+d/2$, $T=2^o\text{C}$, $V/V_C = 2$, $\omega_B = 100\text{Hz}$, $\omega = \omega_A/\omega_B = 1$ and $z \approx 1/2$. The left peak represents the non-translocated events, whereas the other two peaks represent translocation. Inset: The range for which $\lambda$ yields three peaks in $F(t)$ is shown to be $0.1 \leq \lambda \leq 0.3$ when given the above parameters.

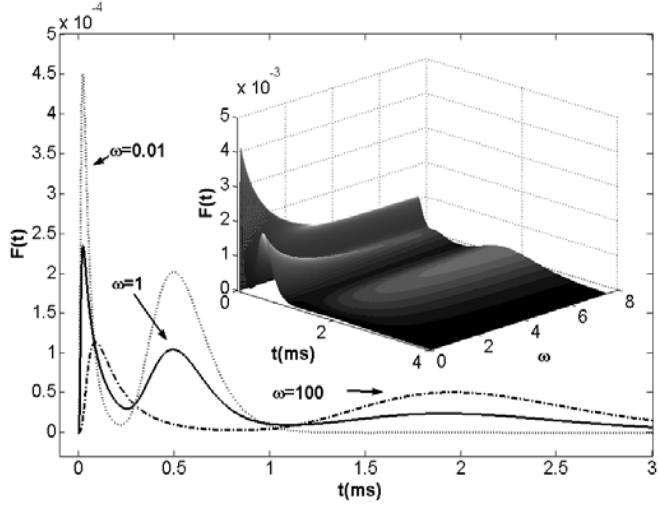

FIG 3 $F(t)$ of poly-$dT_{nu}$ for several values of $\omega_A$ and a fixed $\omega_B$ ($\omega_B = 100$Hz), with $\lambda = 1/4$, and the other parameters as in Fig. 2. Inset: For small values of $\omega$, $F(t)$ displays one translocation peak that corresponds to conformation A whereas for large values of $\omega$, $F(t)$ displays one translocation peak that corresponds to conformation B. For $\omega \approx 1$ two translocation peaks are obtained.

Assuming that the rate of the conformational fluctuations is controlled mainly by temperature, we take $\omega_A$ and $\omega_B$ and as voltage independent in the regime of biological interest $0 \leq V/V_C \leq 3$, using $V_C \approx 50 mV$ [6]. We assume, however, that the result of these fluctuations has weak voltage dependence. Namely, $\lambda(V)$ follows $\lambda(V) \approx \lambda_0 + V/V_\lambda$, which is valid for $V$ such that $\lambda(V) \leq 1$. Here $\lambda_0$ and $V_\lambda$ might be expansion coefficients.

**THE MFPT**

In this section we calculate the MFPT, $\tau$, which allows for an analytical estimation of the characteristic times of the FPT PDF. $\tau$ is calculated by inverting $\mathbf{H}$, which is the matrix on the right hand side of Eq. 2. Using the projection operator technique [14,15] we find

$$\mathbf{H}^{-1} = \begin{pmatrix} \mathbf{A}^{-1}\mathbf{C}(\mathbf{B}-\omega_B) & -\mathbf{A}^{-1}\mathbf{C}\omega_B \\ -\mathbf{A}^{-1}\mathbf{C}\omega_A & \mathbf{A}^{-1}\mathbf{C}(\mathbf{A}-\omega_A) \end{pmatrix} \quad (3)$$

and $\mathbf{C} = (\mathbf{B} - \omega_B - \lambda\omega_A)^{-1}$. After somewhat lengthy calculations, the expression for $\tau$ in the range $V/V_C \geq 1$, and $\omega_A/k, \omega_B/k \ll 1$, reads [13]:

$$\tau \approx \frac{2x\xi_p b^2 d^\mu}{z|q|(1+1/d)} \frac{1}{V-V_C} (P_{A,0} + P_{B,0} \frac{V_\lambda}{V}), \quad (4)$$

where $P_{A,0}$ ($P_{B,0}$) is the probability for the process to start in the A (B) pore conformation. $\tau$ consists of two terms that can be attributed to the A (first term in the brackets) and B (second term in the brackets). When $V_\lambda/V \approx 1$ in the relevant voltages window, the brackets on the right hand side of Eq. 4 can be replaced by 1, indicating that $F(t)$ has one translocation peak that scales close to linearly with $(V-V_C)^{-1}$. As $V_\lambda$ increases, the two terms in Eq. 4 are separated, implying that $F(t)$ has two translocation peaks that are characterized by terms that scale as $(V-V_C)^{-1}$ and $[V(V-V_C)]^{-1}$.

# CONCLUSIONS

The model introduced here describes the translocation of ssDNA through a fluctuating pore structure. As a consequence the ssDNA within the transmembrane pore part is exposed to a changing environment that can lead to three peaks in $F(t)$. This feature is obtained when the following requirements are fulfilled: $\omega \approx 1$, $\omega_B / k \leq 10^{-3}$ and $0.1 \leq \lambda \leq 0.3$. Expanding $\lambda$ up to first order in $V$, the restrictions on $\lambda$ are translated into $V_\lambda$, which is its first order expansion coefficient.

The analytical expression of $\tau$ was shown to scale between a linear to quadratic with $(V/V_C - c)^{-1}$ [$c$ is a constant of the order of *o(1)*], again, sensitive to $V_\lambda$.

Both $F(t)$ and $\tau$ behaviors show that $V_\lambda$ has an important impact on the translocation process and serves as a tuning parameter to the effective dimensionality of the translocation.

The model accounts for the diversity of experimental results discussed in the introduction.

# ACKNOWLEDGMENTS

We acknowledge fruitful discussions with Ralf Metzler and Amit Meller, and the support of the US-Israel Binational Science Foundation and the Tel Aviv University Nanotechnology Center